\documentclass[a4paper]{article}

\def\half{{1\over 2}}

\newcommand{\pdr}{\partial}
\newcommand{\beqs}{\begin{eqnarray}}
\newcommand{\eeqs}{\nonumber\end{eqnarray}}
\newcommand{\beq}{\begin{eqnarray}}
\newcommand{\eeq}{\end{eqnarray}}

\def\m#1{${#1}$}
\def\M#1{\m{#1}}
\def\Em#1{\em  #1}

\def\Lie{{\cal L}}

\def\ben{\begin{enumerate}}
\def\een{\end{enumerate}}
\def\defn{{\bf Defn.}\hfill\linebreak}

\begin{document}

\centerline{\bf Quantization of Contact Manifolds and Thermodynamics}

\begin{center}
{\sf S. G. Rajeev}\\
Department of Physics and Astronomy\\ 
Department of Mathematics\\
University of Rochester\\
  Rochester NY 14627
\end{center}

\centerline{\bf Abstract}

The physical variables of classical thermodynamics occur in conjugate
pairs such as pressure/volume, entropy/temperature, chemical
potential/particle number. Nevertheless, and unlike  in classical
mechanics, there are an odd number of such thermodynamic
co-ordinates. We review the formulation of thermodynamics and
geometrical optics in terms of contact geometry. The Lagrange  bracket provides a generalization of canonical commutation relations. Then we explore the quantization of this algebra  by analogy to the quantization of mechanics. The quantum contact algebra is associative, but the constant functions are not represented by multiples of the identity: a reflection of the classical fact that Lagrange brackets satisfy the Jacobi identity but not the Leibnitz identity for derivations. We verify that this `quantization' describes correctly the passage from geometrical to wave optics as well. As an  example, we work out the  quantum contact geometry  of odd-dimensional  spheres.

\vfill\eject
{\it ``A theory is the more impressive the greater the simplicity of its premises, the more different kinds of things it relates, and the more extended its area of applicability. Therefore the deep impression that classical thermodynamics made upon me. It is the only physical theory of universal content which I am convinced will never be overthrown, within the framework of applicability of its basic concepts.''} -- A. Einstein

\section{Introduction}
\def\Em#1{\em #1}

In classical thermodynamics, as in classical
mechanics, observables come in canonically conjugate pairs: pressure is
conjugate to volume, temperature to entropy, magnetic field to
magnetization, chemical potential to the number of particles etc. An
important difference is that the thermodynamic state space is odd
dimensional.
Instead of the phase space forming a symplectic manifold ( necessarily
even dimensional) the thermodynamic state space is a {\em contact
  manifold}, its odd dimensional analogue\cite{ArnoldGiventhal}. 

Upon passing to the quantum theory, observables of mechanics become
operators; canonically conjugate observables cannot be simultaneously
measured and satisfy the uncertainty principle
\beq
\Delta p\Delta q\geq \hbar.
\eeq
Is there is an analogue to this uncertainty principle\footnote{There
  is
 already an uncertainty relation\cite{Gilmore}  for {\em statistical}
rather then quantum fluctuations of thermodynamica quantities, where
the analogue of \m{\hbar} is \m{kT}.}
  for
thermodynamically conjugate variables? Is there such a thing as
`quantum thermodynamics' where pressure or volume are represented as
operators? 

 The product of thermodynamic conjugates such as 
 \m{\Delta P\Delta V} has the units of energy
rather than action. So if there is an uncertainty relation 
\m{\Delta P\Delta V\geq \hbar_1}, it is clear that \m{\hbar_1} cannot
be Plank's constant as in quantum mechanics.

In principle, there could be macroscopic systems for which the thermal
fluctuations in aggregate quantities such as pressure or volume are
small, yet the quantum fluctuations are not small. For
example\footnote{ I thank Anosh Joseph for discussions on this topic.}, a gas
of cold atoms  confined by a potential has an uncertain  position for
the  `wall'
containing it: the uncertainty in the position of the wall  is of the order of the wavelength \m{\lambda} of the atoms. 
An uncertainty in the position of  the wall by $\Delta x$ leads to an uncertainty in the volume of $\Delta V=A\Delta x$, where \m{A} is the area of the wall.

Similarly, the pressure of
the gas is also uncertain, since pressure is the change of momentum
per unit area per unit time of the particles reflected by the
potential:  the uncertainty of momentum is of order \m{\hbar\over
  \Delta x}. The number of collisions per unit time is \m{\rho v A}
where, \m{v} the typical velocity and
\m{\rho} the number density. Thus we should expect $\Delta P={\hbar\over \Delta x} 
{\rho v A}{1\over A}$
\beq
\Delta P\Delta V\geq \hbar_1=\hbar\nu,\quad \nu \approx  \rho v A.
\eeq
As the system becomes very large, the \m{PV}  scales like the volume
while the r.h.s. of the uncertainty relation  scales as the area; in
this sense it beomes small in many familiar systems.

This justifies the usual practice in quantum statistical mechanics textbooks, of calculating the internal energy  of a quantum system such as a Bose gas or a Fermi gas by averaging over states and then using it in thermodynamics as if it is a classical system to derive the equation of state.

 But there should be systems large enough
for thermodynamics to be applicable but small enough that that the
uncertainty cannot be ignored, of the order of surface effects.
 It would be exciting to test this
prediction experimentally. The appearance of area here is reminiscent
of  the `holographic principle' in quantum gravity. Our  argument suggests that the quantum thermodynamic effects are particularly important 
in systems where the Area and Volume scale the same way, as in hyperbolic spaces.

Another possible application of quantum thermodynamics is to blackholes. Classical
general relativity predicts that blackholes obey the rules of
classical thermodynamics, with the entropy being proportional to the
area of the horizon. Although we don't yet have a quantum theory of
gravity (string theory being the main, but not the only,  candidate)
we can expect that the quantity with the dimensions of energy that
appear in the quantum thermodynamic uncertainty relations is Plank mass. For a black hole that is small enough, it could be important to describe thermodynamic observables such as mass and 
area as operators, thus significantly affecting the debate on the final state of black hole evaporation.

In this paper we will explore the
mathematical  problem of  quantizing contact geometry. This is of
interest for other reasons than thermodynamics in any case.
Quantum or non-commutative notions of geometry are playing an increasing role in many areas of physics and mathematics \cite{Connes}. Deformations of classical notions of Riemannian geometry are central to quantum theories of gravity such as string theory, M-theory and loop quantum gravity. Quantum analogues of symplectic geometry arise as soon one considers the geometry of the phase space in quantum theory. As the odd dmensional sibling of symplectic geometry, contact geometry should also have a quantum analogue, with applications as varied as for contact geometry:
 in constructing knot invariants, or  in quantizing systems with a constraint. Our own original
 motivation was  to understand the possible quantum deformations of the $S^5$ that appear in 
the Maldacena correspondence between gauge and string theories. But we will present the results in a broader context.

We begin with a  review of how contact geometry arises in  classical physics, because much of this material does not seem to be widely accessible in the modern literature in physics. We follow closely   the approach of  Arnold and Giventhal \cite{ArnoldGiventhal} for the classical theory, as a starting point for our quantization. See also the books by Buchdahl \cite{buchdahl} and the influential  review articles by Lieb and Yngvason \cite{LiebYngvason} that revive this classical subject.

\section{Thermodynamics}

Consider a material in an enclosure of volume \m{V}, pressure
\m{P}, temperature \m{T}, entropy \m{S} and
 internal energy \m{U}.
 The first law of thermodynamics says that infinitesimal changes of these  thermodynamic variables  must satisfy 

\begin{eqnarray}
\alpha\equiv dU+PdV-TdS=0
\end{eqnarray}

Although there appears to be  only one condition among the variations of the five
co-ordinates, there is no solution with four independent variables;
i.e.,there is  no 
four-dimensional submanifold  all of whose tangent vectors \m{v}
satisfy the condition \m{i_v\alpha=0}. Of course, this is because the condition above is not a scalar, but  a  `Pfaffian system' of equations. The number of independent variables of such a  system can only be determined by a subtler analysis using the language of differential forms. In fact the theory of differential forms  was orginally developed in part to understand thermodynamics.

If there were four out of five independent variables, there would have been  a function \m{f} that vanished on the hypersurface of solutions.
Also there would be  another function  \m{g} such that \m{\alpha=fdg}.
The integrability condition ( of Frobenius) for that to be the case is \m{\alpha\wedge d\alpha=0}, which is not satisfied in our case. Indeed, even \m{\alpha\wedge(d\alpha)^2\sim dUdPdVdTdS\neq 0} everywhere, so there is not even a three dimensional submanifold that solves   \m{\alpha=0}. 

A submanifold of maximal dimension all of whose tangent vectors are annihilated by \m{\alpha} is called a {\em Lagrangian submanifold}.
In our example, the   dimension of  a  Legendre  submanifold is  two,which therefore is the number of independent thermodynamic degrees of freedom.

In other words,  the first law of thermodynamics implies that only two out of the five variables \m{U,P,V,T,S} are independent: the remaining variables are given by the equations of state. 
The particular two dimensional submanifold chosen as solution will depend on the material since it depends on  the equations of state.  
This key insight is due to J. W. Gibbs\cite{Gibbs}, in his first paper.

So far, the internal energy \m{U} appears to have a 
special status as the `unpaired' variable . However, this is illusory:
the first law of thermodynamics can also be expressed as 
\m{
dS-\beta dU+\tilde  PdV=0
}
where \m{\beta={1\over T},\tilde P={P\over T}} .
This can be viewed as the condition for maximizing \m{S} subject to
the constraint that internal energy \m{U} and volume are held fixed:
then \m{\beta,\tilde P} are the Lagrange multipliers for these constraints.

Giving \m{S} as a function of \m{U,V} ({\em The Fundamental Relation}) is
 a way of determining the Legendre submanifold of a substance: the
 remaining variables are then given as derivatives
\m{
\beta=\left[{\pdr S\over \pdr U}\right]_V,\tilde P=-\left[{\pdr S\over
    \pdr V}\right]_U.
}

 For the monatomic  ideal gas for example,the gas law and
  equipartition of energy give 
 \beq
 PV=nRT ,\quad U={3\over 2}nRT.
 \eeq
 Here \m{n} is the number of atoms in the gas divided by the Avogrado number.
The condition \m{dS-\beta dU+\tilde PdV =0} then determines entropy:
\beq
S=nR\log\left[{U^{3\over 2}\over V}\right]
\eeq
up to a constant.

Conversely, this  Fundamental Relation determines all the other  relations among the thermodynamical quantities.
 The intensive variables \m{\beta,
   \tilde P},  are  given by  derivatives
 w.r.t. their conjugate variables:
\beq
{1\over T}=\left({\pdr S\over \pdr U}\right)_V={3 nR\over 2U},\quad {P\over T}=-\left({\pdr S\over \pdr V}\right)_U={nR\over V}.
\eeq

We can also describe the Legendre submanifold
using  co-ordinates \m{S,P} or \m{T,V} by rewriting the condition
\m{\alpha=0} in the forms
\beq
d[U-PV]+VdP+TdS=0,\quad\ {\rm or}\ d[U+TS]-PdV-SdT=0.
\eeq
We can chose any pair as the fundamental variables as long as they are not conjugate to each other. Each choice provides a co-ordinate system on the Legendre submanifold.

In each picture there is a  {\em Thermodynamical
  Potential}  \m{u}, fundamental variables \m{q^i} and their conjugate
variables \m{p_i}  such  that 
\beq
du-p_idq^i=0.
\eeq
This is the condition for minimizing (or maximizing) \m{u} subject to the
condition that \m{q^i} are held fixed; the conjugate variables \m{p_i}
are  the Lagrange multipliers for these constraints. The  different
equivalent choices of fundamental variables are related by Legendre transformations.


Given a function \m{F} of the five variables and the condition
\M{\alpha\equiv du-p_idq^i=0},
 there is a one-parameter
family of transformations, given as the solutions of the ordinary
differential equations
\beq
\dot q^i={\pdr F\over \pdr p_i},\quad \dot p_i=-{\pdr F\over \pdr
  q^i}-p_i{\pdr F\over \pdr u},\quad \dot 
u=p_i{\pdr F\over \pdr p_i}-F.
\eeq
The tangent to these curves are in the kernel of \m{\alpha}; i.e., any
solution to the above ODEs  will be consistent with the first law of thermodynamics.
  We can  use these transformations to interchange fundamental
variables.

For example, the choice \m{F=\half(p_1^2+q_1^2)}
interchanges \m{q_1\to p_1, p_1\to -q_1} and \m{u\to u-p_1q_1} after a
`time' \m{\pi\over 2}. As another example, \m{p} generates a
scaling of \m{u} and \m{q}.

If the generating function is independent of \m{u}, the transformation
of \m{p,q} are the  canonical transformations familiar from classical mechanics. But in general they are
not. Another, important difference is that   there are here an {\em  odd} number of variables in the thermodynamic  {\em phase space}.

\section{Characteristic Curves}

The same mathematical structures occur other branches of physics, such as classical mechanics and geometrical optics. More generally, in the theory of characteristics of partial differential equations\cite{CourantHilbert}.

Suppose we have a first order quasi-linear PDE
\beq
a^i(q,u){\pdr u\over \pdr q^i}=b(q,u).
\eeq
This is the problem of finding a surface tangential  to the vector
field \m{a^i{\pdr\over \pdr q^i}+b{\pdr \over \pdr u}} at all
points. That is, changing \m{q^i} by \m{a^idt} has the effect of
changing \m{u} by \m{bdt}. In other words, the surface that solves the PDE  is ruled by the integral curves (`characteristic curves') of this vector field.
Hence, solving  the PDE is equivalent to finding the general solution of  the system of ODEs,
\beq
{d q^i\over dt}=a^i(q(t),u(t)),\quad {du\over dt}=b(q(t),u(t)).
\eeq

This idea in fact generalizes even to a nonlinear first order PDE
\beq
F\left(u,{\pdr u\over \pdr q},q\right)=0,
\eeq
 if we allow \m{p_i=
{\pdr u\over \pdr q^i}} as extra variables.
To  the function \m{F(u,p,q)} is associated the ordinary differential equations for characteristic curves
\beq
\dot q={\pdr F\over \pdr p},\quad \dot p=-{\pdr F\over \pdr q}-p{\pdr F\over \pdr u},\quad \dot u=p{\pdr F\over \pdr p}-F.
\eeq

In other words given the condition \m{\alpha\equiv du-pdq=0} on the infinitesimals, and a function \m{F} we can construct the vector field 
\beq
V_F=\left[p{\pdr F\over \pdr p}-F\right]{\pdr \over \pdr u}-\left[{\pdr F\over \pdr q}+p{\pdr F\over \pdr u}\right]{\pdr \over \pdr p}+{\pdr F\over \pdr p}{\pdr \over \pdr q}
\eeq
whose integrals are the characteristic curves.

If we impose the condition 
\beq
du-p_idq^i=0
\eeq
on infinitesimal variations,each  function \m{u(q)} defines a Lagrangean submanifold.
The problem of solving the PDE \m{F(u,{\pdr u\over \pdr q},q)=0}  becomes that of finding an intersection of a  Lagrangean submanifold with the hypersurface defined by \m{F(u,p,q)=0} . So every hypersurface on a contact manifold must be `ruled' by curves, the characteristic curves of the corresponding PDE.

In a medium whose  refractive index depends on position and direction, the eikonal equation of geometrical optics takes the form
\beq
g^{ij}(x){\pdr u\over \pdr x^i}{\pdr u\over \pdr x^j}=k^2
\eeq
where \m{k} is the wave number.
Thus,
\beq
F(u,p,x)=g^{ij}(x)p_ip_j-k^2.
\eeq
The characteristic curves are the light rays; they are the solutions\cite{buchdahl}  of the ODE above with this choice of \m{F}.
\section{Classical Mechanics }

Although it has become fashionable to formulate classical mechanical in terms of symplectic geometry,
the Hamilton-Jacobi formulation is best understood in terms of an odd-dimensional phase space of co-ordinates
\m{(u,q,p)} where \m{u} is the eikonal or
 Hamilton's principal function\footnote{ For uniformity of notation with the last section we will denote the eikonal by \m{u} rather than \m{S}  as is common in mechanics textbooks.
}. The condition \m{du-p_idq^i=0}  is satisfied if 
\beq
p_i={\pdr u\over \pdr q^i}.
\eeq

 The constancy of energy
\beq
F(q,p)=H(q,p)-E=0
\eeq
defines a hypersurface on this odd-dimensional phase space which is ruled by the characteristic curves generated by this function
\beq
\dot q={\pdr H\over \pdr p},\quad \dot p=-{\pdr H\over \pdr q},\quad \dot u=p{\pdr H\over \pdr p}-H+E.
\eeq 
The first pair are the Hamilton's equations; the last equation gives the variation of the eikonal along a classical trajectory.

This larger phase space allows as symmetries the Legendre transformations whose generator can depend on the eikonal \m{u} in addition to \m{p} and \m{q}. As we saw, these are more general than the above canonical transformations whose generators only depend on \m{p} and \m{q}. Thus they can accommodate the general first order partial differential equation rather than just the Hamilton-Jacobi equations above.

The first order  PDE of mechanics and optics are  in fact,
the approximation to the wave equation in the limit of small wavelength.
Thus it becomes interesting to develop a unifying framework for
quantizing contact structures that include mechanics, optics and thermodynamics. There are also interesting applications to fluid mechanics \cite{Norbury}.

\section{Legendrian Knots}
 We digress briefly to make contact with a fashionable topic of mathematics\cite{thurston}.
An embedding  of  $S^1$ in $R^3=\{(u,q,p)\}$ is a  {\em  Legendrian knot }  if all its  tangent vectors are  in the kernel of the contact form \m{du-pdq} . 
Any closed curve in the plane with \m{\oint
 pdq=0} will give a knot, simply by choosing \m{u} to be integral of \m{pdq} along the curve. Conversely, every 
Legendrian knot is determined by its projection to the \m{(p,q)} plane. A continuous one parameter family  of contact transformations that does not make the curve intersect itself is a {\em contact isotopy}. Invariants of Legendrian knots under such contact isotopies are of interest in topology of three manifolds\cite{thurston}
 .  Since the condition of being a contact isotopy is stronger than an isotopy, it is knots that are equivalent in the usual sense might be  distinct as Legendrian knots: there are more invariants in  Legendrian knot theory. Ideas from quantum theory have re-invigorated conventional knot theory \cite{witten}; perhaps a theory of quantum contact manifolds can do the same for Legendrian knots.

\section{Contact Structure}
 It is time to reformulate the above nineteenth century physics in the language of twentieth century mathematics.

\defn
A {\em contact form} on a manifold of dimension $2n+1$ is a one-form that satisfies
\beq
\alpha\wedge (d\alpha)^n\neq 0
\eeq
at each point.
Two such forms are considered equivalent if they only differ by multiplication by a positive function (a `gauge transformation'):
\beq
\alpha\sim f\alpha.
\eeq
A {\em  contact manifold} has a  contact form in each co-ordinate patch, with such  positive functions  relating  overlapping patches.

A {\em contact structure} should be thought of as the equation \m{\alpha=0} which picks out a subspace of the tangent space at each point of the manifold. However, the 
Frobenius integrability condition for these to fit together as tangent spaces of some submanifold is maximally violated.

A submanifold  all of whose  the tangent vectors will satisfy
\m{\alpha=0} is said to be `integral'. If the contact manifold is of
dimension \m{2n+1}, the largest dimension for an integral submanifold
will be \m{n}. Such a maximal integral submanifold is called a {\em
  Legendre submanifold}.


 An analogue of Darboux's theorem says that there is a local co-ordinate system ( and choice of gauge) in which
contact form is
\beq
\alpha=du-\sum_{i=1}^np_idq^i.
\eeq
Thus $R^{2n+1}$ with this choice is the basic example of a contact
manifold. A Legendre submanifold is given by the \m{n}
 dimensional subspace with co-ordinates \m{p^i}, holding \m{u,q^i}
 constant. Another Legendre submanifold is given by \m{(u(q),{\pdr
     u\over \pdr q},q)} for some generic function \m{u(q)}.

The condition for a vector field \m{V} to preserve a contact structure is  that there exist a function \m{g_V} such that 
\beq
\Lie_V\alpha=g_V\alpha.
\eeq
The commutator of two such functions will still satisfy this condition.

If we choose local co-cordinates such that \m{\alpha=du-p_idq^i},such a vector field is always of the form
\beq
V=\left[p{\pdr F\over \pdr p}-F\right]{\pdr \over \pdr u}-\left[{\pdr F\over \pdr q}+p{\pdr F\over \pdr u}\right]{\pdr \over \pdr p}+{\pdr F\over \pdr p}{\pdr \over \pdr q}
\eeq
where the {\em  hamiltonian}  of \m{V} is \m{F=-i_V\alpha} and \m{g_V=-{\pdr F\over \pdr u}}.

There is thus a one-one correspondence between contact vector fields and their
hamiltonians. Unlike for symplectic vector fields, a  constant added to
\m{F} {\em  will} change \m{V_F}. 
Note that the integral curves of the contact vector field are precisely 
the characteristic curves of \m{F}  we
obtained earlier. Since \m{V_F(F)=-F{\pdr F\over \pdr u}}, if the
initial value of \m{F} is zero it will remain zero for ever. Thus the
solution to \m{F(u,{\pdr u\over \pdr q},q)=0}, given the value of \m{u} on a boundary curve in the \m{q}-space, can be obtained by
starting at each point on the boundary and evolving along the integral curves of
\m{V_F}. 

If we replace $F$ by a function $\phi(F)$ (where $\phi:R\to R$ has non-zero derivative everywhere) the vector field changes as $V_F\mapsto \phi'(F)V_F$: the integral curves are unchanged except for their parametrization. Thus each hypersurface in a contact manifold determines a family of characteristic curves that lie on them. Through each point on the hypersurface passes exactly one such curve: these curves {\em  rule} the hypersurface.

It will be of interest to look at the special case \m{\Lie_V\alpha=0}.
Such vector fields preserve the volume form \m{\alpha(d\alpha)^n} and
so might also be called {\em incompressible} contact vector
fields. Clearly they are determined by a generating function that is
independent of \m{u}, basically canonical transformations.

\subsection{Digression: Reeb Dynamics}

A contact form by itself also  defines a family of curves.
 The two-form \m{d\alpha} is of maximal rank;
so there is a vector field, unique up to multilication by a non-zero
function, such that 
\m{
i_Vd\alpha=0.
}
Its integral curves are then well defined: a change \m{V\to fV} only
affects the parametrization of the curves. This is the Reeb dynamics of a contact form \cite{Blair}. A contact manifold 
does not however determine a contact form: the choice of a particular
contact form representing  a contact structure is analogous to the 
choice of a generating function $F$  in the last section. We have to  choose here a section of a line bundle; i.e., make  a choice of gauge representing the contact form.
We can see that these Reeb curves extermize the action \m{\int_\gamma
  \alpha}. 

This dynamics is {\em  not} invariant under the transformation \m{\alpha\to
  f\alpha} but is instead invariant under the addition of an exact form
\m{\alpha\to \alpha +d\lambda}.

If \m{\alpha=Hdt-pdq} with \m{H} depending on \m{t,p,q} describes a
time dependent hamiltonian system.  The action principles states that
 the dynamics consist of curves that extremize  the action \m{\int[pdq-Hdt]}; i.e., satisfy the
Hamilton-Jacobi equations. Thus time dependent hamiltonian mechanics
is exactly the  Reeb dynamics for the above contact form. This is a
different point of view on mechanics from  the one using contact
structures.
Important unsolved problems in this subject include the Weinstein
conjecture on the existence of a periodic orbit for the dynamics on
any compact contact manifold\cite{weinstein}.

\section{Canonical Quantization in Contact Geometry}

 The standard example of a symplectic manifold is a co-tangent
  bundle; the analoguous example  for a contact manifold is the {\em  projective}
  co-tangent bundle. In  co-ordinates \m{x^\mu} on a space \m{X}, the
  co-tangent bundle  \m{T^*X} has a
 one-form \m{\theta=k_\mu dx^\mu} and symplectic form \m{dk_\mu dx^\mu}.  On the
 fibers we have an action of the 
group \m{R^\times} of non-zero real numbers, \m{k_\mu\mapsto \lambda
  k_\mu}. The quotient of \m{T^*X-X}
 under this action of \m{R^\times} is the projective co-tangent bundle \m{PT^*X}.

Under this action,
\m{\theta\mapsto \lambda \theta}, so it becomes  a one-form \m{\alpha}
(defined only up to multiplication by a non-zero scalar function) on
the quotient. This is the 
 contact structure on \m{PT^*X}.

If \m{\dim X=n+1} the contact manifold is of dimension \m{2n+1}. The
real projective spaces
 which are the fibers over each point of $X$ are Legendre submanifolds. A hypersurface in this contact manifold can be given by an equation 
\beq
F(x,k)=0
\eeq
where $F:T^*X\to R$ is a homogenous function on the cotanget space:
\beq
F(x,\lambda k)=\lambda^rF(x,k).
\eeq
These curves will rule the hypersurface $F(u,q,p)=0.$
Such a hypersurface will define a family of characteristic curves that rule it. It will be convenient to make a choice of this function that has as simple a dependence on $k$ as possible (e.g., polynomial if possible) because this will simplify the quantization. 

This point of view is particularly appropriate  for null geodesics on a pseudo-Riemannian manifold $X$. The phase space for these geodesics is the projective co-tangent space. There is a  natural  hypersurface $g^{\mu\nu}k_\mu k_\nu=0$ on this contact manifold. The characteristic curves of this hypersurface are the null geodesics.
A `quantization' should recover the wave equation on \m{X}.

Whenever a classical dynamical system can be formulated in terms of the characteristic curves of a hypersurface $F(x,k)=0$  on a projective cotangent bundle $PT^*X$, we have a quantization in the `Schrodinger picture'. The wave functions of the system are then
complex valued functions $\psi:X\to C$ satisfying the wave equation
\beq
\left[F\left(x,{\pdr\over \pdr x}\right)+ \cdots \right]\psi=0.
\eeq
The $\cdots$ denotes lower derivative terms that represent the ordering ambiguities: the rule of canonical quantization can only give the highest order piece as only the principal symbol is known in the classical theory. 
Remember that for the equation $F(x,k)=0$ to be well-defined on the projective space, $F(x,k)$ must be homogenous in $k$. Thus the constant $-i\hbar$ in the usual rule for canonical quantization $k\mapsto -i\hbar {\pdr\over \pdr x}$ drops out of the wave equation.

In the eikonal approximation $\psi(x)=e^{{i\over \hbar}u(x)}$ with small $\hbar$ we will get back the equation for the Legendre submanifold determined by $F$:
\beq
F\left(x,{\pdr u\over \pdr x}\right)=0.
\eeq

Applied to the  case of  null geodesics this gives the wave equation
\beq
\left[g^{\mu\nu} {\pdr\over \pdr x^\mu}{\pdr\over \pdr x^\nu}+\cdots\right]\psi=0.
\eeq  
Physically, this is not quantization as such, but the passage from geometrical optics to wave optics.

If we determine the lower order terms by the requirement of invariance under co-ordinate transformations, we get  the d'Alembertian operator
\beq
 {\pdr\over \pdr x^\mu}\left(\surd[-g] g^{\mu\nu}{\pdr\psi\over \pdr x^\nu}\right)=0
\eeq

When there is no natural identification of the contact manifold as a projective cotangent space, we need a more abstract approach.

\section{The Lagrange Bracket}
Let us recall again  familiar notions from classical mechanics.

A {\em  symplectic form} is a closed two-form that is non-degenerate:
\beq
d\omega=0,\quad i_v\omega=0\Rightarrow v=0.
\eeq
This of course requires the manifold carrying $\omega$ to be even dimensional.

A symplectomorphism (`canonical transformation') is a diffeomorphism that preserves the sympletic form, $\phi^*\omega=\omega$. Infinitesimally, a symplectic vector field satisfies $\Lie_v\omega=0$.
Since $\Lie_v\omega=d(i_v\omega)$, this implies that locally there is a function (`generating function') such that 
\beq
i_v\omega=dg_v.
\eeq
The commutator of two symplectic vector fields is also symplectic. This defines a commutator (`Poisson bracket') on the generating functions:
\beq
\{g_1,g_2\}=r(dg_1,dg_2)
\eeq
where $r$ is the inverse tensor of $\omega$. These brackets satisfy the axioms of a Poisson Algebra.

\defn

A {\em  Poisson Algebra}  is a commutative algebra $A$ with identity along with a bilinear $\{,\}:A\otimes A\to A$  that satisfies 

\ben
\item \m{\{g_1,g_2\}=-\{g_2,g_2\}} 
\item  \m{\{\{g_1,g_2\}, g_3\}+\{\{g_2,g_3\}, g_1\}+\{\{g_3,g_1\}, g_2\}=0}\quad {\rm Jacobi identity}
\item \m{\{g_1,g_2g_3\}=\{g_1,g_2\}g_3+g_2\{g_1,g_3\}}\quad {\rm Leibnitz Rule}
\een
The last property implies that the Poisson bracket of a constant with any function is zero.

The basic example is the algebra of functions on the plane:
\beq
\{g_1,g_2\}={\pdr_x g_1}\pdr_yg_2-\pdr_yg_1\pdr_x g_2.
\eeq

More generally the set of function on a symplectic manifold form a Poisson algebra, with bracket we gave earlier. Conversely, if the Poisson algebra is non-degenerate (the only elements that have zero Poisson bracket with everything are constants) it arises from a symplectic manifold this way.

Also, the set  of functions on the dual of a Lie algebra is a Poisson algebra (Kirillov):
\beq
\{F,G\}(\xi)=i_\xi[dF,dG]
\eeq

The commutator of two contact vector fields is again a contact vector field. This
induces a bracket on functions, called the {\em  Lagrange bracket}
\footnote{ We follow the terminology of Arnold\cite{ArnoldGiventhal}. 
There are also some other unrelated things  called Lagrange brackets
in for example, the textbook by Goldstein.}:
\beq
(F,G)=F{G_ u}-{F_u}G+p\left(F_{p}G_u-F_uG_{p}\right)+F_pG_q-F_qG_p
\eeq
with the summations over indices  implied.

This satisfies the Jacobi identity but not the Leibnitz rule:  even the bracket of the constant with a function may not be zero:
\beq
(1,G)=G_u.
\eeq 

{\bf Defn}\hfill\linebreak
A commutative algebra ${\cal A}$ with identity and a bilinear $(,):{\cal A}\times {\cal A}\to {\cal A}$  is a {\em Generalized Poisson Algebra} if
\begin{enumerate}
\item \m{(g_1,g_2)=-(g_2,g_2)} 
\item   \m{((g_1,g_2), g_3)+((g_2,g_3), g_1)+((g_3,g_1), g_2)=0}
\item \m{(g_1,g_2g_3)=(g_1,g_2)g_3+g_2(g_1,g_3)+(1,g_1)g_2g_3.\quad}  {\rm The Generalized Leibnitz Rule}
\end{enumerate}

The main point is that the constant function is  no longer in the center. The commutant of the constant function is a Poisson algebra.


With the choice $\alpha=du-pdq$ on the simplest case of $R^3$, the  analogue of the canonical commutation relations can now be worked out:
\beqs\nonumber
(p,q)=w,\quad &&(w,q)=0,\quad  (w,p)=0\\\nonumber
(u,q)=0,\quad && (u,p)=p,\quad (u,w)=w
\eeqs
The element $w$ (which is simply  the constant function \m{1}) is not central anymore. \m{w,p,q} span a Heisenberg sub-algebra. \m{u} generates the  automorphism of the Heisenberg algebra which scales $w,p$. Since the Leibnitz rule is replaced by the new identity,  we have to be careful about using these commutation relations to derive brackets for more general functions. e.g., \m{(u,pq)=0}.

\section{ Deformation Quantization of Contact Structures}

Recall that  given a constant Poisson tensor \m{r^{ij}}  on a vector space, we can quantize
it by defining the star product on functions
\beq
f*g(\xi)=\left[ e^{-{i\hbar\over 2}{\pdr\over \pdr \xi^i}r^{ij} {\pdr\over\pdr \xi^{j'}}  }f(\xi)g(\xi')
\right]_{\xi=\xi'}
\eeq

If we expand in powers of \m{\hbar}, to zeroth order we get just the
pointwise product, then the Poisson bracket then various higher order
derivatives that fit amazingly into an associative product when all
terms are takem into account.

 We will now present a similar way of
turning the space of functions on a contact vector space into an
associative algebra: the zeroth order will be the pointwise product,
the first order the Lagrange bracket and so on.

The key idea is to note that a contact vector space \m{V} is always the
`projectivization'  of some symplectic vector space \m{\tilde V}: the
functions on \m{V} lift uniquely  to homogenous functions of degree one on
\m{\tilde V}. Contact vector fields lift to vector fields on \m{\tilde
  V} that commute with the scaling. So we can lift them up, multiply
them and then project to get the star product on the contact vector space.


Given a function on \m{V}, we define a function on \m{\tilde V},
\beq
\tilde F(w,u,q,p)=wF(u,q,wp).
\eeq
Then
\beq
\left[{\pdr \tilde F\over  \pdr w}\right]_{w=1}=F(u,q,p)+p{\pdr F \over \pdr p}
\eeq
so that 
\beq
\{\tilde F,\tilde G\}_{w=1}=(F,G).
\eeq
We use this idea to extend the \m{*}-product as well:

\beq
F*G=
\left[ e^{{-{i\hbar\over 2}\left({\pdr\over \pdr w}{\pdr\over \pdr
          u'}+{\pdr\over \pdr q}{\pdr\over \pdr p'}
-{\pdr\over \pdr w'}{\pdr\over \pdr u}-{\pdr\over \pdr q'}{\pdr\over \pdr p}
\right)}} wF(u,q,wp)w'G(u',q',w'p')
\right]_{w=w'=1,u=u',p=p',q=q'}.
\eeq
It is clear that this multiplication is  associative as it is  a special case of the usual $*$-product of Moyal. Moreover, to leading order in $\hbar$, it is the commutative product plus the Lagrange bracket:
\beq
F*G=FG-{i\hbar\over 2}(F,G)+\cdots
\eeq
This non-commutative multiplication now contains all the quantum effects.
By expanding the exponential in a power series we can get a more explicit form of the $*$-product.


It is convenient to think of this algebra as represented on wave
functions that depend on \m{u,q}. The {\em  correspondence principle} is
\beq
F(u,q,p)\mapsto \hat F=F\left(u,q,(-i\hbar)^2{\pdr^2\over \pdr q\pdr u}
\right)\left[-i\hbar{\pdr\over \pdr u}\right]+\cdots
\eeq
the dots being terms that needed to be added to make the operator hermitean.
The precise formula is
\beqs
&&\hat F \psi(q,u)=\cr
&&\int wF\left({u+u'\over 2},{q+q'\over 2},wp\right)e^{{i\over \hbar}[p\cdot(q-q')+w(u-u')]}\psi(q',u')
{dpdvdq'du'\over (2\pi)^{n+1}}
\eeqs

Even the constant function is represented by an operator: \m{1\mapsto
  {-i\hbar}{\pdr \over \pdr u}}.
 For example, the  hypersurface  \m{p^2+V(q)-E=0},  corresponding to a hamitlonian that is independent of \m{u}, leads to the Schr\"odinger-like equation
\beq
\left\{\left[(-i\hbar)^2{\pdr^2\over \pdr q\pdr u}\right]^2+V(q)-E\right\}
\left[-i\hbar{\pdr\over \pdr u}\right]\psi(q,u)=0
\eeq
The ansatz \m{\psi(q,u)=e^{{i\over \hbar} u}\psi(q)} then yields the usual Schr\"odinger equation
 \beq
\left\{\left[(-i\hbar){\pdr\over \pdr q}\right]^2+V(q)-E\right\}\psi(q)=0
\eeq 
 for \m{\psi(q)}.

More generally, given a function \m{F(u,q,p)} that does depend on \m{u}, classically we have first order PDE
\beq
F\left(\chi(q),q,{\pdr \chi\over \pdr q}\right)=0
\eeq
which defines a Legendre submanifold. Upon quantization we get the {\em
linear} differential equation
\m{
\hat F\psi(q,u)=0
}
where \m{\hat F} is given by the integral formula above. Up to ordering ambiguities
\beq
F\left(u,q,(-i\hbar)^2{\pdr^2\over \pdr u\pdr q}\right)
\left\{(-i\hbar){\pdr \over \pdr u}
\right\}\psi(q,u)=0
\eeq

In the semi-classical approximation the solution of the above wave equation  is 
\beq
\psi(q,u)\approx e^{{i\over\hbar}[u+\chi(q)]}
\eeq with \m{\chi(q)} satisfying the first order PDE
\beq
F\left(\chi(q),q,{\pdr \chi\over \pdr q}\right)=0
\eeq

\section{ Odd-dimensional Quantum Spheres}

There are several examples of quantum spheres known in the literature \cite{conneslandi}. They are  quantum deformations of the sphere thought of as the set of unit vectors in Euclidean space. If this vector space is even dimensional, it can carry a symplectic structure and then the odd dimensional sphere embedded in it inherits a contact structure. We will study the quantum deformation of these `contact spheres' as an example.

Let us begin with a constant symplectic structure $\omega{ij}$ on a vector space of dimension $2n$ and a sphere $n^in^i=1$ in it.  This is just the energy surface of a harmonic oscillator; there is a vector field on the surface of the sphere which defines the time evolution of the hamiltonian function  $x^ix^i$ on the ambient vector space containing the sphere. If the characteristic values of the anti-symmetric matrix (which are proportional to the periods of the normal modes of the harmonic oscillator) are all equal, this  co-ordinate will be periodic, the `angle' variable conjugate to the hamiltonian $x^ix^i$.

Given a function on the sphere we can expand it in `spherical harmonics':
\beq
F(n)=F_\emptyset+F_in^i+F_{ij}{n^in^j\over 2!}+\cdots  .
\eeq
The coefficients are symmetric traceless tensors.
We can lift this function on the sphere to the ambient vector space
\beq
\tilde F=x^2\left[F_\emptyset+F_ix^i+F_{ij}{x^ix^j\over 2!}+\cdots
\right]
\eeq
The  factor $x^2$  in front ensures that the Poisson bracket on the ambient space restricts to the Lagrange bracket on the sphere:
\beq
\left\{\tilde F,\tilde G\right\}_{x^2=1}=(F,G).
\eeq

Quantization of the algebra of functions on the contact manifolds is now just the quantization of the lift of these functions. More precisely we already have the usual Moyal $*$-product on the function on the  symplectic vector space. We define the star product of functions on the sphere to be
\beq
F*G=\left[\tilde F*\tilde G \right]_{x^2=1}.
\eeq
Thus even the constant function goes over to an operator that is {\em not} the multiple of the identity.

We can also represent this algebra on the Hilbert space of the harmonic oscillator. The metric $\delta_{ij}$ and the symplectic structure together define a complex structure on the  vector space. The quadratic function $x^2$ being the  hamiltonian is then the function 
\beq
H(z)=\sum_{a=1}^n\omega_az^a z^{\bar a}.
\eeq
Here $\omega_i$ are the frequencies of the normal modes of the harmomic oscillator. Or, $\omega_a$ are equal to half the reciprocal of the characteristic values of the symplectic tensor.
Every  polynomial on the vector space can be written as a polynomial in $z^a,z{\bar a}$:
\beq
\Phi= \Phi_\emptyset+ \Phi_az^a+\Phi_{\bar a}z^{\bar a}+\Phi_{ab}{1\over 2!}z^az^b+\Phi_{a{\bar a}}z^az^{\bar a}+\Phi_{{\bar a}{\bar b}}{1\over 2!}z^{\bar a}z^{\bar b}+\cdots
\eeq

The Hilbert space of the harmonic oscillator is the space of polynomials in the $z^a$. The inner product is defined by declaring the 
monomials
\beq
z_1^{N_1}\cdots z_n^{N_n}
\eeq
to be orthonormal. The above polynomial becomes the operator 
\beq
\hat \Phi= \Phi_\emptyset+ \Phi_aA^{\dag ^a}+\Phi_{\bar a}A^{\bar a}+\Phi_{ab}{1\over 2!}A^{\dag a}A^{\dag b}+\Phi_{a{\bar a}}{1\over 2!}(A^{\dag a}A^{\bar a}+A^{\bar a}A^{\dag a})+\Phi_{{\bar a}{\bar b}}{1\over 2!}A^{\bar a}A^{\bar b}+\cdots
\eeq
We have chosen the symmetric ordering of operators.

Now we can see how each function on the sphere goes over to an operator on this quantum Hilbert space. The constant function $1$ goes over to the hamiltonian operator:
\beq
\hat H=\sum_a\omega_a [A^{\dag a}A^{\bar a}+\half].
\eeq
More generally the function $F(n)=F_\emptyset+F_in^i+F_{ij}{n^in^j\over 2!}+\cdots $ on the sphere becomes
\beq
\hat F=F_\emptyset \hat H+F_a{1\over 2}[\hat H, A^{\dag a}]_+\cdots
\eeq

\section{Acknowledgement}
This work was supported in part by the Department of Energy   under the  contract number 
 DE-FG02-91ER40685. I also acknowledge discussions with A. P. Balachandran, B. Khesin, G. Landi, V. P. Nair and especially with Anosh Joseph.  Special thanks also to Klaus Bering for a careful reading of an earlier version of this paper.

\end{document}